\documentstyle[12pt]{article}

\begin{document}

\title{Gravitational Collapse and Cosmic Censorship}
\author{Robert M. Wald\\
         {\it Enrico Fermi Institute and Department of Physics}\\
         {\it University of Chicago}\\
         {\it 5640 S. Ellis Avenue}\\
         {\it Chicago, Illinois 60637-1433}}
\maketitle

\begin{abstract}

We review the status of the weak cosmic censorship conjecture, which
asserts, in essence, that all singularities of gravitational collapse
are hidden within black holes.  Although little progress has been made
toward a general proof (or disproof) of this conjecture, there has
been some notable recent progress in the study of some examples and
special cases related to the conjecture. These results support the
view that naked singularities cannot arise generically.

\end{abstract}
\newpage

\section{Introduction}
\label{int}

It has long been known that under a wide variety of circumstances,
solutions to Einstein's equation with physically reasonable matter
must develop singularities \cite{he}. In particular, if a sufficiently
large amount of mass is contained in a sufficiently small region,
trapped surfaces must form \cite{sy} or future light cone
reconvergence should occur \cite{p1}, in which case gravitational
collapse to a singularity must result. One of the key outstanding
issues in classical general relativity is the determination of the
nature of the singularities that result from gravitational collapse.

A key aspect of this issue is whether the singularities produced by
gravitational collapse must always be hidden in a black hole---so that
no ``naked singularities'', visible to a distant observer, can ever
occur. The conjecture that, generically, the singularities of
gravitational collapse are contained in black holes is known as the
{\it weak cosmic censorship conjecture} \cite{p2}. (A much more
precise statement of this conjecture will be given in Sec.~\ref{for}
below.) A closely related conjecture, known as the {\it strong cosmic
censorship conjecture} \cite{p3}, asserts that, generically, timelike
singularities never occur, so that even an observer who falls into a
black hole will never ``see'' the singularity. This paper will focus
exclusively on the weak cosmic censorship conjecture.

In the past seven years, there have been two articles in the New York
Times indicating the demise of weak cosmic censorship \cite{nyt}. The
main theme of this paper is that such reports of the death of cosmic
censorship are greatly exaggerated: Although very little progress has
been made toward a general proof of the weak cosmic censorship
conjecture, it remains in good health, and, indeed, is probably
healthier today than at any time in the past.

\section{Formulation of the weak cosmic censorship conjecture}
\label{for}

The issue at the heart of the weak cosmic censorship conjecture can be
expressed in graphic terms by posing the following question: Could a
mad scientist---with arbitrarily large but finite resources---destroy
the universe? We know that, in principle, such a mad scientist could
produce a spacetime singularity by gathering together a sufficiently
large amount of mass into a sufficiently small region. The essential
content of the weak cosmic censorship conjecture is the assertion
that, if he were to do so, neither the singularity he would produce
nor any of its effects could ever propagate in such a way as to reach
a distant observer. Of course, even if weak cosmic censorship fails,
the universe might still be protected from destruction by mad
scientists, since even if naked singularities were produced, they
might always be of a benign character, with well defined
rules---presumably arising from quantum gravity---governing dynamical
evolution in their presence. However, if weak cosmic censorship fails,
then in a literal sense, we would come face-to-face with the laws of
quantum gravity whenever gravitational collapse to a naked singularity
occurs in distant regions of our universe.

To formulate the statement of the weak cosmic censorship conjecture
more precisely, we first need to make precise the notion of the
``finiteness'' of the resources available to our mad scientist. This
notion is well modeled by restricting consideration to spacetimes
containing an asymptotically flat initial data surface, i.e., a
hypersurface $\Sigma$ on which the induced spatial metric, $h_{ab}$
becomes Euclidean at asymptotically large distances from some compact
(``central'') region, and the extrinsic curvature, $K_{ab}$, of
$\Sigma$ goes to zero at a suitable rate at infinity. If matter fields
are present, additional asymptotic fall-off conditions for the matter
fields also would be required. (The precise asymptotic conditions on
this initial data that would be most suitable to impose probably would
best be left open until further progress is made in investigations of
cosmic censorship, and will not be considered here.) In particular, it
should be noted that the restriction to asymptotically flat initial
data ensures that our mad scientist initially has only a finite total
amount of energy at his disposal, but it does not place any direct
restrictions on the initial conditions he might set up in the
``central region'' of the spacetime.

The basic idea of the weak cosmic censorship conjecture is that,
starting from these initial conditions, any sufficiently distant
observer will neither encounter any singularities nor any effects
arising from---i.e., propagating out of---singularities. To make this
idea more precise, we need a suitable notion of a spacetime being
asymptotically flat ``at large distances and at late times''. Such a
notion is provided by the requirement that the spacetime be
asymptotically flat at future null infinity. The standard definition
of asymptotic flatness at future null infinity requires that one be
able to conformally embed the spacetime in a suitable way into a
spacetime with a boundary, ${\cal I}^+$, which, roughly speaking,
provides endpoints for the null geodesics which propagate to
asymptotically large distances. The precise details of this definition
are not crucial to our discussion and can be found in standard
references (see, e.g., \cite{he}, \cite{w1}). The precise smoothness
requirements most suitable to impose at ${\cal I}^+$ undoubtedly will
depend on the precise choice of asymptotic conditions on the initial
data (see above), and will not be considered here.

For a spacetime, $(M, g_{ab})$, which is asymptotically flat at future
null infinity, the {\it black hole} region, ${\cal B}$, of the
spacetime is defined by
\begin{equation}
{\cal B} = M - I^-({\cal I}^+)
\label{bh}
\end{equation}
where the chronological past, $I^-$, is taken in the (``unphysical'')
conformally completed spacetime. The {\it event horizon}, ${\cal H}$,
of the black hole is defined to be the boundary of ${\cal B}$ in $M$
\begin{equation}
{\cal H} = \partial {\cal B}.
\label{eh}
\end{equation}

We are now in a position to give a relatively precise formulation of
weak cosmic censorship:

\medskip

\noindent {\bf Weak cosmic censorship conjecture}: Let $\Sigma$ be a
3-manifold which, topologically, is the connected sum of ${\cal R}^3$
and a compact manifold. Let $(h_{ab}, K_{ab}, \psi)$ be nonsingular,
asymptotically flat initial data on $\Sigma$ for a solution to
Einstein's equation with {\it suitable} matter (where $\psi$ denotes
the appropriate initial data for the matter). Then, {\it generically},
the maximal Cauchy evolution of this data is a spacetime, $(M,
g_{ab})$ which is asymptotically flat at future null infinity, with
complete ${\cal I^+}$.

\medskip

In this formulation of the conjecture, the asymptotic flatness of $(M,
g_{ab})$ (with complete ${\cal I^+}$) ensures that sufficiently
distant observers can live out their lives in their entirety, free
from the effects of any catastrophic events occurring in the central
region of the spacetime. Furthermore, the fact that these observers
lie in the domain of dependence of $\Sigma$ implies that they also are
free from any non-deterministic effects that might occur if
singularities are produced. If singularities are produced, they cannot
be seen from ${\cal I^+}$.

The above conjecture remains somewhat imprecise on account of the two
words written in italics. In order for the matter to be ``suitable'',
it clearly is necessary that the coupled Einstein-matter field
equations have a well posed initial value formulation. It undoubtedly
also should be required that the matter stress-energy tensor satisfy
suitable energy conditions, such as the dominant energy condition. In
addition, it would seem reasonable to require that the matter be such
that, in any fixed, globally hyperbolic, background spacetime (such as
Minkowski spacetime), one always obtains globally nonsingular
solutions of the matter field equations starting from regular initial
data; otherwise, any ``naked singularities'' produced in the dynamical
evolution of the Einstein-matter equations may have nothing to do with
gravitational collapse. Note that this latter condition would rule out
fluids as ``suitable matter'' (in particular, on account of ``shell
crossing'' singularities and shocks), although fluid examples remain
quite valuable as simple matter models for testing behavior related to
cosmic censorship (see, e.g., \cite{jd1}). It is not clear whether any
further restrictions should be imposed---or, indeed, if some of the
above restrictions should be weakened somewhat. In any case, the
``suitable'' matter fields certainly should include the Maxwell field
and the Klein-Gordon scalar field.

The ``generic'' condition was inserted in the above conjecture because
it would not be fatal to the physical content of the conjecture if
examples exist where dynamical evolution produces naked singularities,
provided that the initial data required for these examples is so
special that it would be physically impossible to achieve. A way to
express the idea that no generic violations of weak cosmic censorship
occur would be to require that all initial data giving rise to
violations of the behavior specified in the conjecture is confined to
a ``set of measure zero'' and/or a ``set whose closure has empty
interior''. Unfortunately, it is far from clear precisely what measure
or topology should be imposed on the space of initial
data. Undoubtedly, it will be necesary to develop a much deeper
insight into the dynamics implied by Einstein's equation before a
natural choice of measure or topology will emerge, and I feel that the
precise definition of ``generic'' would best be left open until that
point.

Does the weak cosmic censorship conjecture hold? To answer this
question, we would need to know a great deal about the global
properties of solutions to Einstein's equation. Global existence of
solutions with nearly flat initial data has been proven
\cite{ck}. Thus, weak cosmic censorship holds for nearly flat
data---where the nonlinear effects of general relativity are too weak
to produce any singularities at all. However, mathematical techniques
have not progressed to a stage where a direct attempt at a general
proof of the weak cosmic censorship conjecture would be
feasible. Thus, the evidence both for and against the validity of weak
cosmic censorship has been largely of an ``anecdotal'' or
``circumstantial'' nature, or has been confined to some very
restricted cases (such as spherical symmetry). In the remaining
sections of this paper, I shall briefly summarize much of this
evidence.

\section{Some evidence in favor of cosmic censorship}
\label{evi}

In this section, I shall describe some analyses in support of cosmic
censorship. Most of the basic ideas described here are at least 20 years
old, but some notable recent progress has occurred.

\subsection{Stability of black holes}
\label{sta}

If weak cosmic censorship fails, then gravitational collapse can
(generically) result in a naked singularity rather than a black
hole. If so, then it is quite possible that the formation of a black
hole would be a non-generic outcome of collapse. In that case, one
might expect to see evidence of this in linear perturbation theory off
of a background spacetime containing a black hole. Specifically, one
might expect an initial, smooth perturbation to grow without bound on
the black hole horizon, signaling the conversion of the black hole
into a naked singularity. (In linear perturbation theory, a blow-up of
the perturbation on the horizon could occur only at asymptotically
late times, but it could occur at a finite time in the nonlinear
theory.) Thus, the study of linear perturbations off of a black hole
background provides an excellent testing ground for weak cosmic
censorship. A demonstration of the linear instability of black holes
would effectively disprove weak cosmic censorship, whereas a
demonstration of their linear stability would provide some notable
evidence in support of it.

The first analysis of the linear stability of the Schwarzschild black
hole was given in 1970 by C.V. Vishveshwara \cite{v}, who established
its stability to axial (i.e., ``odd parity'') perturbations. Shortly
thereafter, a convincing demonstration of the stability of the
Schwarzschild black hole---together with detailed information about
the decay properties of the perturbations---was given by Price
\cite{pr}. More recently, a completely rigorous proof of the
boundedness of perturbations at asymptotically late times has been
given \cite{kw}.

The analysis of the stability of a Kerr black hole is much less
tractable than the Schwarzschild case. Nevertheless, Whiting \cite{wh}
succeeded in proving that no unstable modes exist. Thus, it appears
that weak cosmic censorship has passed the crucial test of stability
of black holes to general, linear perturbations.

Another test of stability can be performed for extremal charged Kerr
black holes, i.e., charged Kerr black holes whose mass, $M$, angular
momentum, $J$, and charge, $Q$, satisfy
\begin{equation}
M^2 = Q^2 + (J/M)^2
\label{ex}
\end{equation}
No stationary black hole solutions exist when the right side of
Eq.~(\ref{ex}) exceeds the left side. Thus, if one can get an extremal
black hole to ``swallow'' an object whose charge and/or angular
momentum is sufficiently large compared with its mass, there would be
no black hole final state available for the system to settle down to.
Presumably, a naked singularity then would result. However, an
analysis of test particle motion in an extremal charged Kerr
background indicates that it is not possible to get a black hole to
swallow too much charge or angular momentum \cite{w2}. Interestingly,
a similar anaylsis of test particle motion for certain extremal
``black hole'' solutions of the Einstein-Maxwell equations with a
positive cosmological constant---where the spacetime is asymptotically
DeSitter rather than asymptotically flat---indicates that
``over-charging'' can occur in this case \cite{bhkt}. However, these
``black holes'' are only rough analogs of black holes in
asymptotically flat spacetimes; in particular, in these solutions
there are naked singularities as seen from the DeSitter analog of
${\cal I}^+$. Even if these ``black holes'' could be ``destroyed'' by
perturbations, this would not contradict the formulation of weak
cosmic censorship given above for asymptotically flat
spacetimes---though it would provide evidence against some
formulations of strong cosmic censorship.

\subsection{Failed counterexamples}
\label{fai}

A class of possible counterexamples to weak cosmic censorship
involving collapsing shells of null dust was proposed by Penrose
\cite{p4} and generalized by Gibbons \cite{g1} over 25 years ago. A
similar class of possible counterexamples also can be given for time
symmetric initial data. In the analysis of these classes of possible
counterexamples, a key role is played by the following two results
from the theory of black holes in general relativity (see, e.g.,
\cite{he},\cite{w1}):

\begin{itemize}
\item If weak cosmic censorship holds, then every trapped surface must
be entirely contained within a black hole. (Here, a {\it trapped
surface}, $S$, is a compact, 2-dimensional surface having the property
that the convergence of both the outgoing and ingoing null geodesics
normal to $S$ is everywhere positive.)

\item If weak cosmic censorship holds and if matter satisfies the null
energy condition (i.e., if $T_{ab} k^a k^b \geq 0$ for all null $k^a$),
then the area of the event horizon of a black hole cannot decrease
with time.

\end{itemize}

Now consider a convex shell of null dust with flat interior, which
collapses from infinity down to an infinite density singularity. (Note
that such null dust presumably would not qualify as ``suitable
matter'' in our formulation of the weak cosmic censorship conjecture,
but if the inequality given below could be violated for null dust, it
presumbly also could be violated for ``suitable matter''.) In this
example, there are two free functions of two variables which may be
specified arbitrarily, characterizing the initial ``shape'' and
initial mass density of the shell. Except in the case of a spherical
shell with constant mass density, the solution exterior to the shell
is not known. Nevertheless, enough information about the solution can
be deduced to make an interesting test of cosmic censorhip.
Specifically, the Bondi mass, $M$, infinitesimally outside of the
shell at past null infinity can be computed, and it seems reasonable
to expect that there will exist exterior solutions with total ADM mass
equal to $M$ (or, at least, arbitrarily close to $M$). Furthermore, by
integrating the Raychaudhuri equation for outgoing null geodesic
congruences across the delta-function mass distribution on the null
shell one can determine the presence of trapped surfaces lying
infinitesimally outside of the shell.

Suppose, now, that a trapped surface, $S$, lies infinitesimally
outside of the shell, and let $A(S)$ denote its area. As mentioned
above, if weak cosmic censorship holds, then $S$ must lie within a
black hole. Let $A_0$ denote the area of the 2-surface obtained by
intersecting the event horizon, ${\cal H}$, of this black hole with
the null shell. Then, assuming the validity of weak cosmic censorship,
the following string of inequalities should hold:
\begin{equation}
A(S) \leq A_0 \leq 16 \pi M^2_{\rm bh} \leq 16 \pi M^2
\label{PG}
\end{equation}
Here, the first inequality follows from the fact that $S$ lies within
the black hole and the shell is infalling. The second inequality
follows from the area nondecrease theorem (see above) together with
the fact that the maximum possible area of a stationary black hole of
mass $M_{\rm bh}$ is achieved by the Schwarschild value, $16 \pi
M^2_{\rm bh}$. (Here it is assumed that the black hole settles down to
a stationary final state.)  The final inequality expresses the fact
that the mass of the final black hole cannot exceed the total ADM mass
of the spacetime.

Remarkably, the inequality $A(S) \leq 16 \pi M^2$ implied by
Eq.~(\ref{PG}) involves only quantities which can be computed without
knowing the solution exterior to the shell, and thus can be readily
checked. Failure of this inequality to hold in any example would be
nearly fatal to cosmic censorship, as only a few small loopholes would
remain---such as the possible ``unsuitablity'' of the null dust
matter, the possibly ``non-generic'' nature of the example, and the
(very remote) possiblity that the black hole does not become
asymptotically stationary.

By making use of the fact that the interior of the shell is flat, the
issue of whether the above inequality holds can be reduced to the
issue of whether a certain ``isoperimetric inequality'' holds for
(topological) spheres of non-negative mean curvature embedded in
Euclidean 3-space. Until very recently, the inequality $A(S) \leq 16
\pi M^2$ had been proven only in some special cases. However, recent
results of Trudinger \cite{t} on strengthened isoperimetric
inequalities has enabled a general proof to be given that $A(S) \leq
16 \pi M^2$ in all cases; see \cite{g2} for further discussion.

A similar argument---showing that an analog of (\ref{PG}) must hold if
weak cosmic censorship is valid---can be given for the case of time
symmetric initial data (i.e., $K_{ab} = 0$) on a spacelike
hypersurface $\Sigma$. A {\it minimal surface}, $S$, on $\Sigma$ is a
compact (without boundary) 2-surface on which $p \equiv h^{ab} p_{ab}
= 0$ everywhere, where $p_{ab}$ is the extrinsic curvature of $S$ in
$\Sigma$.  In the case where $K_{ab} = 0$, both sets of null geodesics
orthogonal to a minimal surface $S$ will have vanishing expansion, so
$S$ will be marginally trapped. It then follows that $S$ must lie
within a black hole. Now, let $S_{\rm out}$ be the outermost minimal
surface on $\Sigma$. Since the intersection of the black hole horizon,
${\cal H}$, with $\Sigma$ must lie outside of (or coincide with)
$S_{\rm out}$, and since $S_{\rm out}$ is the outermost minimal
surface, it follows that the area, $A_0$, of ${\cal H} \cap \Sigma$
cannot be smaller than $A(S_{\rm out})$. Hence, in analogy with
Eq.~(\ref{PG}), we obtain
\begin{equation}
A(S_{\rm out}) \leq A_0 \leq 16 \pi M^2
\label{JW}
\end{equation}
where $M$ denotes the ADM mass of the spacetime. Again, both $A(S_{\rm
out})$ and $M$ can be calculated directly from the initial data given
on $\Sigma$, without the necessity of evolving the data off of
$\Sigma$. In essence, Eq.~(\ref{JW}) shows that if cosmic censorship
is valid, then a strengthened version of the positive mass theorem
must hold for time symmetric initial data whenever minimal surfaces
are present. 

An argument for the validity of Eq.~(\ref{JW}) was given in \cite{jw},
but it relied on the assumed existence of particular foliation of
$\Sigma$ by 2-surfaces, as first proposed by Geroch \cite{ge} in an
argument for the positivity of total mass. However, very recently, a proof
of existence of the required foliation has been given \cite{hi}, thus
establishing that Eq.~(\ref{JW}) does, indeed, hold for general, time
symmetric initial data.

Of course, the proof of the inequalities (\ref{PG}) and (\ref{JW}) for
the very restricted classes of spacetimes to which they apply is a far cry
from even the beginnings of a general proof of weak cosmic
censorship. Nevertheless, it is very hard to understand {\it why}
these highly nontrivial inequalities should hold unless weak cosmic
censorship can be thought of as providing the underlying physical
reason behind them. Thus, while it is not clear how much ``objective
evidence'' in favor of cosmic censorship is provided by the failure of
these counterexamples, their failure has given many researchers
considerable confidence in the validity of weak cosmic censorship.

\section{The hoop conjecture}
\label{hoo}

Although the results of the previous section provide some suggestive
evidence in favor of cosmic censorship, it is not surprising that not
all researchers have been convinced by this evidence, and some serious
doubts about the general validity of weak cosmic censorship have been
expressed. Many of these doubts have centered upon an idea known as
the {\it hoop conjecture}, which has been formulated as follows (see
\cite{th1} or box 32.3 of \cite{mtw}):

\medskip

\noindent {\bf Hoop conjecture}: Black holes with horizons form when
and only when a mass $M$ gets compacted into a region whose
circumference in EVERY direction is ${\cal C} \leq 4 \pi M$.

\medskip

\noindent (Here, one envisions ``passing a hoop'' of circumference $4 \pi M$
around the matter in every direction to test this criterion.) The
basic idea intended to be expressed by this conjecture is that
gravitational collapse in all three spatial dimensions must occur in
order for a black hole to form. If collapse occurs in fewer dimensions
(i.e., to a 2-dimensional ``pancake'' configuration or a 1-dimensional
``spindle'' configuration), then the hoop conjecture is normally
interpreted as asserting that a naked singularity should result.

Clearly, the above statement of the hoop conjecture in not intended to
be mathematically precise, and there are some obvious mathematical
difficulties with its formulation. Probably the most serious of these
difficulties arises from the fact that the ``only when'' portion of
the conjecture cannot be expected hold unless ``gravitational energy''
is included in the ``mass $M$'', since there should be no difficulty
forming a black hole out of a sufficient concentration of
gravitational radiation. However, there is no local notion of
``gravitational energy'' in general relativity. In addition, there is
no obvious notion of the ``circumference'' a world tube in spacetime
(such as the world tube containing the ``mass $M$'' of the
conjecture), since arbitrarily near (in spacetime) to any given
2-dimensional surface exterior to a world tube are 2-surfaces
(approximated by suitably chosen broken null surfaces) with arbitraily
small circumference in every direction. In other words, one can pass
an arbitrarily small hoop around {\it any} concentration of mass by
making segments of the hoop move in a suitable, relativistic manner as
one moves the hoop around the mass. In order for the conjecture to
have meaning, it is necessary to specify the choice of spacelike
slicing on which the circumference is to be measured, but it is not
obvious how this should be done.

I do not know how to overcome the difficulties in the formulation of
the ``only when'' portion of the hoop conjecture. In the theorem of
Schoen and Yau \cite{sy}---establishing the existence of trapped (or
anti-trapped) surfaces in certain situations when a sufficiently large
amount of matter is compacted into a sufficiently small region---the
difficulties in defining the ``circumference'' are avoided by using
the internal geometry on a spacelike slice to measure the ``size'' of
the region. (This measure of the size also can be made arbitrarily
small by appropriate choices of slicing, but the matter
energy-momentum density on the slice is then necessarily affected in a
corresponding manner.) It should be noted that the theorem of Schoen
and Yau does not actually establish a version of the ``when'' half of
the hoop conjecture unless one {\it assumes} weak cosmic censorship,
since the presence of a trapped surface necessarily implies the
existence of a black hole only under that assumption (see the
beginning of subsection \ref{fai}).

The main motivation for the ``when'' half of the hoop conjecture
appears to have arisen from the study of the collapse of matter with
cylindrical symmetry (i.e., rotational symmetry about an axis together
with translational symmetry along that axis). It has long been known
that cylindrically symmetric fluids can collapse to singularities, but
no trapped surfaces ever form \cite{th1}, \cite{ch}, and the
singularities are ``visible'' from infinity. These examples do not
directly provide counterexamples to weak cosmic censorship because the
cylindrically symmetric spacetimes are not asymptotically flat in the
required sense: The matter distribution and curvature extend to
infinity along the axis, and even in the directions perpendicular to
the axis, the metric approaches flatness too slowly. However, these
examples might seem to suggest that weak cosmic censorship could be
violated in the collapse of a very long but finite ``spindle'' of
matter.

However, although cylindrically symmetric fluids can collapse to a
singularity, it does not appear that such singularities occur
generically or that they occur at all with ``suitable
matter''. Specifically, for a dust cylindrical shell, an arbitrarily
small amount of rotation causes a ``bounce'' \cite{at}.  Furthermore,
it has been shown that in the vacuum and Einstein-Maxwell cases, no
singularities whatsoever occur in cylindrical collapse \cite{bcm}.
Thus, even if it turns out that the collapse of a finite ``spindle''
of fluid matter can produce naked singularities similar to those which
occur in the exactly cylindrical case, it seems unlikely that spindle
collapse will produce naked singularities that satisfy either the
``suitable matter'' or ``generic'' provisions of the weak cosmic
censorship conjecture.

Nevertheless, Shapiro and Teukolsky \cite{st} performed numerical
calculations of the collapse of highly prolate gas spheroids and found
behavior which they interpreted as both supporting the hoop conjecture
and providing likely counterexamples to the weak cosmic censor
conjecture. In their calculations, they evolved a number of spacetimes
describing collapsing gas spheriods using maximal time slicing. They
found that when the spheroid was highly prolate, a singularity formed
just exterior to the ends of the spheroid---at which point, of course,
they could no longer continue their numerical evolution. They then
searched for trapped surfaces on the maximal hypersurfaces and found
that none were present. They interpreted this as indicating that the
singularity might be naked.

As already indicated above, even if the singularity they found were
shown to be naked, it would be far from clear that the ``suitable
matter'' and ``generic'' provisions of the weak cosmic censor
conjecture could be satisfied. However, it also should be emphasized
that the absence of a trapped surface lying on their maximal slices in
the portion of the spacetime that they constructed does not provide
much evidence that the singularity they found is naked. Indeed, even
in Schwarzschild spacetime, it is possible to choose a (highly
non-spherically-symmetric) time slice which comes arbitrarily close to
the singularity inside the black hole, and yet has no trapped surfaces
contained within its past \cite{iw}. A good test of whether something
like this might be occurring in the Shapiro-Teukolsky examples is
provided by the Penrose-Gibbons spacetimes described above in
subsection~\ref{fai}. One can choose the shell of null dust to be
highly prolate and one can arrange the collapse of the shell so
that---in the flat hyperplane slicing of the flat interior of the
shell (the analog of maximal slicing in the Shapiro-Teukolsky
examples)---the singularity occurs at the ends first. One may then
analyze whether trapped surfaces occur just outside the null shell
and, if so, where they are located. Examples can be given where
trapped surfaces exist, but no trapped surface is wholly contained to
the past of the ``last hyperplane'' prior to encountering the
singularity \cite{to}. These examples suggest that if a significantly
larger portion of the Shapiro-Teukolsky spacetimes were constructed
(which could be done by using different choices of time slicing), then
trapped surfaces enclosing the singularity might be found. Thus, at
present, there do not appear to be strong reasons to believe that the
singularities found by Shapiro and Teukolsky are actually naked.

In summary, if weak cosmic censorhip holds, then the Schoen and Yau
theorem \cite{sy} can be viewed as giving a precise statement and
proof of a version of the ``when'' portion of the hoop
conjecture---although, as pointed out to me by N. O'Murchadha, the
applicability of this theorem is extremely restrictive. However,
formidable difficulties would have to be overcome even to give a
precise formulation of the ``only when'' portion of the hoop
conjecture, since one would need a notion of the gravitational
contribution to the ``mass $M$''. Furthermore, even if some version of
the ``only when'' portion of the conjecture turns out to be valid,
there need not be any conflict with weak cosmic censorship, since it
could well be the case that if suitable matter fails to be
sufficiently compacted in all three spatial directions, then,
generically, no singularity forms. In any case, I am not aware of any
results related to the hoop conjecture which cast a serious doubt on
the validity of weak cosmic censorship.

\section{The spherically symmetric Einstein-Klein-Gordon System}
\label{ekg}

As previously emphasized, a general analysis of the validity of cosmic
censorship would appear to require a much greater mastery of the
global properties of solutions to Einstein's equation than is
presently achievable. Therefore, it is natural to focus attention on
more tractable special cases. The assumption of spherical symmetry
greatly simplifies the analysis of gravitational collapse, and,
largely for that reason, has been widely studied. It should be kept in
mind that there is no guarantee that phenomena found in spherically
symmetric gravitational collapse are representative of general
phenomena. In particular, a phenomenon which is generic under the
restriction to spherical symmetry need not be generic when that
restriction is removed. Nevertheless, it is instructive to explore the
phenomena that occur in spherical gravitational collapse, and to
determine to what degree cosmic censorship holds in that case.

By Birkhoff's theorem the gravitational field itself has no dynamical
degrees of freedom in the spherical case, so it is essential to have
matter degrees of freedom present. Spherical collapse has been most
widely studied with fluid matter, particularly ``dust'', i.e, a
perfect fluid with $P=0$. Numerous examples have been found where
naked singularities occur (see, e.g., \cite{ysm}, \cite{c0}). However,
most of these examples appear to be non-generic in character, and all
of them appear to rely on properties of fluid matter that also would
allow one to produce singularities during evolution in flat spacetime.
In particular, the ``shell crossing'' and ``shell focusing'' naked
singularities found in the collapse of dust matter appear to depend
crucially on the ability to ``aim'' the dust so as to produce infinite
density before the self-gravitation of the dust becomes large. Thus,
in order to obtain more insight into the validity of cosmic censorship
in the spherically symmetric case, it would appear to be necessary to
study examples with a more ``suitable'' form of matter.

A suitable form of matter which provides an excellent testing ground
for cosmic censorship is provided by a massless Klein-Gordon scalar
field, $\phi$. The complete system of equations for the
Einstein-Klein-Gordon system are
\begin{equation}
\nabla^a \nabla_a \phi = 0
\label{KG}
\end{equation}
\begin{equation}
G_{ab} = 8 \pi [\nabla_a \phi \nabla_b \phi -
\frac{1}{2}g_{ab} \nabla_c \phi \nabla_c \phi]
\label{E}
\end{equation}
When restricted to spherical symmetry, Eqs.(\ref{KG}) and (\ref{E})
simplify greatly from the general case, but, as will be indicated
further below, they still provide a rich dynamics.

In a series of papers, Christodoulou \cite{c1}-\cite{c5} has given a
remarkably complete analysis of the singularities that can arise in
spherically symmetric solutions to the Einstein-Klein-Gordon
equations. Christodoulou considered evolution from initial data posed
on a future null cone, $C^+_0$, with vertex on the world line $\Gamma$
corresponding to the center of spherical symmetry, $r = 0$. (Here $r$
denotes the usual Schwarzschild radial coordinate defined by $4 \pi
r^2 = A$, where $A$ denotes the area of the orbit the rotation group.)
The initial data on $C^+_0$ can be characterized by the function
\begin{equation}
\alpha \equiv \frac{d}{dr} (r \phi)
\label{alpha}
\end{equation}
which may be freely specified on $C^+_0$ subject to asymptotic
conditions and boundary conditions at $r = 0$. In \cite{c3}
Christodoulou showed that unique solutions of bounded variation
(defined precisely in \cite{c3}) exist provided that the initial data
is such that the function $\alpha$ is of bounded variation on $C^+_0$.

In \cite{c4} Christodoulou investigated the global behavior spherically
symmetric Einstein-Klein-Gordon solutions which possess the following
additional conformal symmetry: There exists a one-parameter group of
diffeomorphisms (parametrized by $\lambda$) under which, for some
constant $k$,
\begin{equation}
g_{ab} \rightarrow \lambda^2 g_{ab}, \,\,\,\, r \rightarrow \lambda r,
\,\,\,\, \phi \rightarrow \phi - k \ln \lambda
\label{con}
\end{equation}
Such solutions will not be asymptotically flat, but the initial data
for these solutions can be suitably ``truncated'' so as to yield
asymptotically flat data, and consequently they are relevant for
testing cosmic censorship. By analyzing these solutions, Christodoulou
proved \cite{c4} that {\em there exist choices of asymptotically flat
initial data which evolve to solutions with a naked singularity}. In
these solutions, a singularity first forms at the origin and then
propagates out to infinity along a (singular) future null cone,
reaching ${\cal I}^+$ at a finite retarded time. Thus, ${\cal I}^+$ is
incomplete in the maximally evolved spacetime. The null cone
singularity is of a rather mild type, in that the curvature remains
bounded as one approaches the cone away from the vertex, although
derivatives of the curvature blow up on the cone. Christodoulou also
proved that there exist choices of asymptotically flat initial data
which evolve to what he referred to as ``collapsed cone
singularities''. The solutions with collapsed cone singularities can
be thought of as describing black holes of vanishing mass which
possess a singular event horizon. In these solutions, ${\cal I}^+$ is
complete---so the conditions of our formulation of the weak cosmic
censorship conjecture are satisfied---but the singularity is not
really hidden in a black hole and from infinity one can ``see'' events
which are arbitrarily close to the singularity.\footnote{It should be
noted that all of Christodoulou's examples of naked singularities
arise from initial data on $C^+_0$ of a low differentiability class
(but, of course, with $\alpha$ of bounded variation, so that the
initial value formulation is well posed). His examples with collapsed
cone singularities can have initial data of arbitrarily high (but
finite) differentiability.}

Thus, the analysis of \cite{c4} established for the first time that
naked singularities can arise by the gravitational collapse of
``suitable matter''.\footnote{The existence of a naked singularity had
previously been claimed for a particular scale invariant solution
of the Einstein-Klein-Gordon equations (i.e., a solution satisfying
Eq.~(\ref{con}) with $k = 0$), which has been studied by a number of
authors \cite{mr} (see also \cite{c3}). However, the analysis given in
\cite{c4} shows that this solution (or, more precisely, suitable
asymptotically flat ``truncations'' of this solution) actually corresponds to a
collapsed cone singularity rather than a naked singularity.} A great
deal of insight into the circumstances under which naked singularities
are produced was provided by Choptuik's \cite{cho} numerical
investigations of the behavior of Einstein-Klein-Gordon solutions
which are ``just on the verge'' of collapsing to a black
hole. Choptuik considered various one-parameter families of initial
data with the property that for small values of the parameter, the
incoming scalar waves are weak and disperse back to infinity, whereas
for large values of the parameter, the incoming scalar waves are
strong and collapse to a Schwarzschild black hole. Choptuik then tuned
the parameter to the ``borderline value'' where collapse first occurs,
and used mesh refinement techniques to study the properties of the
``borderline solution'' near $r = 0$ in detail. Remarkably, he found
that for all of the one-parameter families he considered, the
borderline solution always asymptotically approached a particular
``universal solution''. Furthermore, this universal solution was found
to possess a discrete self-similarity, i.e., it admits a
diffeomorphism (as opposed to a one-parameter group of diffeomorphisms)
satisfying Eq.~(\ref{con}) with $k = 0$. Neither the universality of
the borderline collapse behavior nor the discrete self-similarity of
the universal solution had been anticipated prior to Choptuik's
analysis. Most importantly for the issue of cosmic censorship, the
numerical investigations by Choptuik (confirmed by others \cite{hs})
indicated that the borderline solutions possess naked singularities of
a nature similar to Christodoulou's examples.

Similar discrete or continuous self-similarity has been found to occur
in the borderline solutions for a number of other systems (see, in
particular, \cite{ss}). Presumably, these borderline solutions also
possess naked singularities (although I am not aware of demonstrations
of this). However, in some systems where there exist unstable,
stationary, nonsingular solutions---in particular, in the
Einstein-Yang-Mills system---some borderline solutions approach one of
these stationary, nonsingular solutions rather than a self-similar
solution \cite{ccb}.

The fact that naked singularities in the spherically symmetric
Einstein-Klein-Gordon system were encountered in Choptuik's numerical
calculations only for the borderline solutions suggests that the
occurrence of naked singularities is non-generic. Clearly, no
definitive conclusions in this regard can be drawn from numerical
studies. However, an analytic demonstration of the non-generic
character of naked singularities has recently been given by
Christodoulou \cite{c5}. In order to state this result, it is useful
to classify solutions arising from the maximal evolution of
asymptotically flat initial data on $C^+_0$ as follows:

\begin{itemize}

\item[] {\bf case (i)}: No singularities at all occur at any finite
advanced time and ${\cal I}^+$ is future complete.\footnote{In this
case, the spacetime also must be future timelike and null geodesically
complete, with the possible exception of the ``central geodesic''
$\Gamma$ at $r = 0$, whose completeness was not explored by
Christodoulou.} This case necessarily arises when the initial data is
sufficiently ``small'' \cite{c3}.

\item[] {\bf case (ii)}: A singularity forms at a finite advanced time,
but it is entirely contained within a black hole, as in the ``standard
picture'' of gravitational collapse. In particular, ${\cal I}^+$ is
complete, and in order to reach the singularity, an observer must
pass through a (non-singular) event horizon.

\item[] {\bf case (iii)}: Neither case (i) nor case (ii) holds.

\end{itemize}

\noindent Note that under this classification, case (iii) includes all
solutions with naked singularities and collapsed cone singularities,
as well as any solutions with other, as yet undiscovered, pathologies
that might be viewed as contrary to the spirit of the weak cosmic
censorship conjecture.

Christodoulou proved the following \cite{c5}: {\em Consider any
initial data---characterized by the function $\alpha_0$---which
evolves to a spacetime in category (iii) above. Then there exists a
continuous function $f$ such that for any real number $c \neq 0$, the
initial data characterized by $\alpha = \alpha_0 + cf$ evolves to a
spacetime in category (ii).}\footnote{The borderline solutions studied
by Choptuik presumably comprise a surface, ${\cal S}$, of co-dimension
$1$ in the space of initial data. In order to have the above property
at $\alpha_0 \in {\cal S}$, Christodoulou's one parameter family must
be tangent to ${\cal S}$, and ${\cal S}$ must be convex (as viewed
from the ``black hole side'') in the direction defined by this
family.} In other words, by an arbitrarily small perturbation of the
initial data, a spacetime containing a naked singularity (or a
collapsed cone singularity or other pathology) can be converted to a
black hole. Thus, within the class of initial data of bounded
variation, solutions with naked singularities are non-generic in the
above, precise sense.

Although restricted to the case of the spherically symmetric
Einstein-Klein-Gordon equations, the above result provides the first true
cosmic censorship theorem for a nontrivial system. 

\section{Conclusions}

Although the question of whether weak cosmic censorship holds remains
very far from being settled, there appears to be growing evidence in
support of its validity. This evidence consists primarily of the
stability of black holes (see subsection \ref{sta}), the proof of the
failure of certain classes of counterexamples (see subsection
\ref{fai}), and the proof of a cosmic censorship theorem for the
spherically symmetric Einstein-Klein-Gordon system (see
Sec.~\ref{ekg}).

\bigskip

\noindent {\bf Acknowledgements}

I wish to thank Demetrios Christodoulou for reading the
manuscript. This research was supported in part by NSF grant PHY
95-14726 to the University of Chicago.

\end{document}